\documentclass[11pt]{article}
\usepackage{erice,epsfig}
\usepackage{subfigure}

\def\Journal#1#2#3#4{{#1} {\bf #2}, #3 (#4)}

\newcommand\EPC{{Eur. Phys. J.} C}
\newcommand\NIMA{Nucl. Inst. and Meth. A}
\newcommand\NPS{Nucl. Phys. B (Proc. Suppl.)}
\newcommand\NPB{ Nucl. Phys. B}
\newcommand\PLB{Phys. Lett.  B}
\newcommand\PRL{Phys. Rev. Lett.}
\newcommand\PRD{Phys. Rev. D}

\newcommand\ea{{\em et al.}}

\newcommand{\amu}[1][]{\ensuremath{a_{\mu^{#1}}}}
\newcommand{\gm}{\ensuremath{(g-2)}}
\newcommand{\wa}{\mbox{\ensuremath{\omega_a}}}

\newcommand{\wpt}{\mbox{\ensuremath{\widetilde{\omega}_p}}}

\begin{document}
\vspace*{4cm}
\title{THE BNL MUON ANOMALOUS MAGNETIC MOMENT MEASUREMENT}

\author{ DAVID W. HERTZOG }

\address{Department of Physics, University of Illinois at Urbana-Champaign \\
1110 W. Green St., Urbana, IL 61801, USA}

\author{Representing the E821 Muon g-2 Collaboration~\cite{collab} }
\address{ }
\maketitle\abstracts{ The E821 experiment at Brookhaven National
Laboratory is designed to measure the muon magnetic anomaly,
$a_{\mu}$, to an ultimate precision of 0.4~parts per million
(ppm). Because theory can predict $a_{\mu}$ to 0.6~ppm, and
ongoing efforts aim to reduce this uncertainty, the comparison
represents an important and sensitive test of new physics. At the
time of this Workshop, the reported experimental result from the
1999 running period achieved $a_{\mu^+}~=
11\,659\,202(14)(6)\times10^{-10}$ (1.3~ppm) and differed from the
most precise theory evaluation by 2.6 standard deviations.
Considerable additional data has already been obtained in 2000 and
2001 and the analysis of this data is proceeding well. Intense
theoretical activity has also taken place ranging from suggestions
of the new physics which could account for the deviation to
careful re-examination of the standard model contributions
themselves.  Recently, a re-evaluation of the pion pole
contribution to the hadronic light-by-light process exposed a sign
error in earlier studies used in the standard theory. With this
correction incorporated, experiment and theory disagree by a
modest 1.6 standard deviations.  }

\section{Introduction}
\begin{figure}
\begin{center}
\subfigure[QED]{\psfig{figure=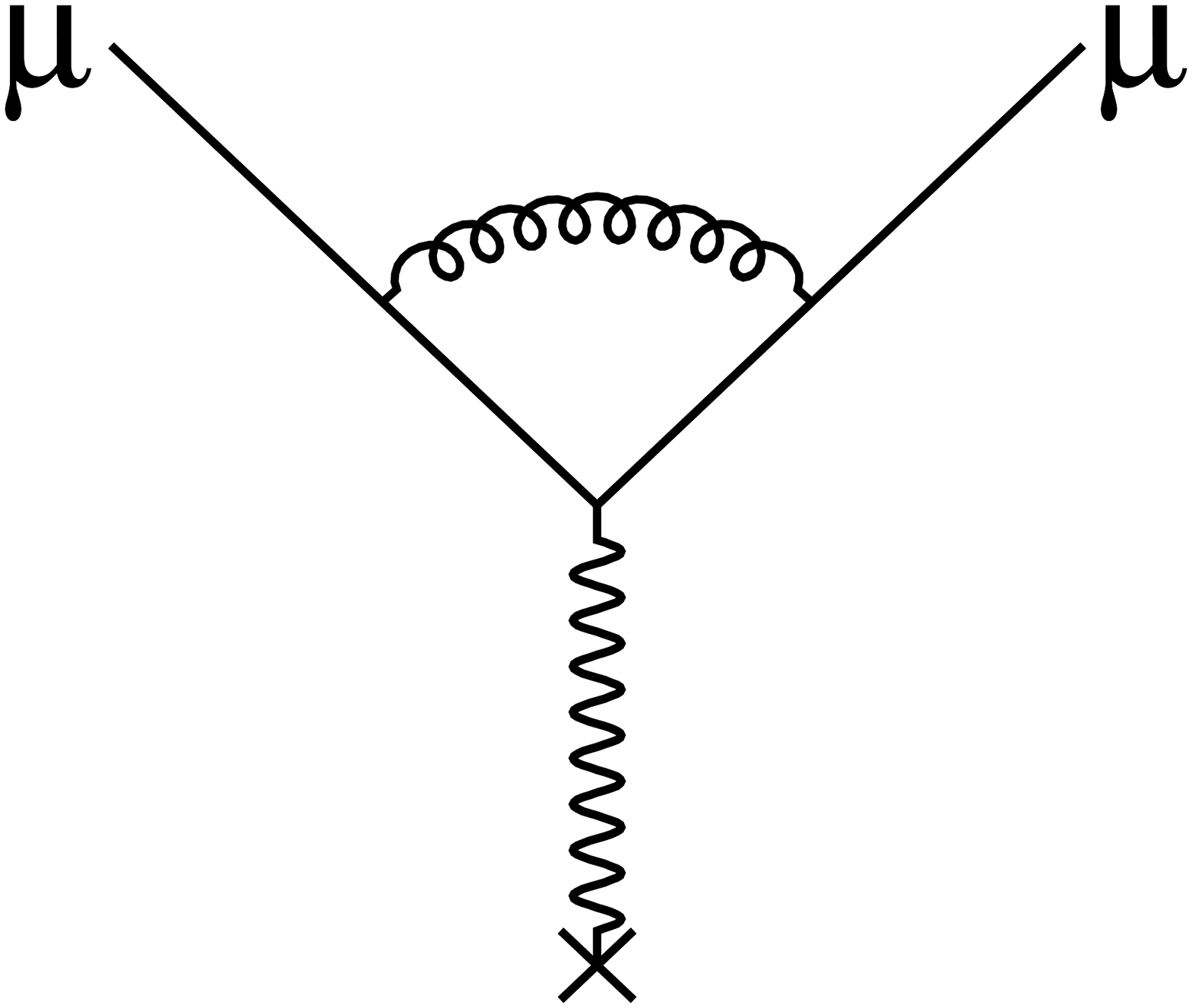,width=.24\textwidth}}
\subfigure[Weak]{\psfig{figure=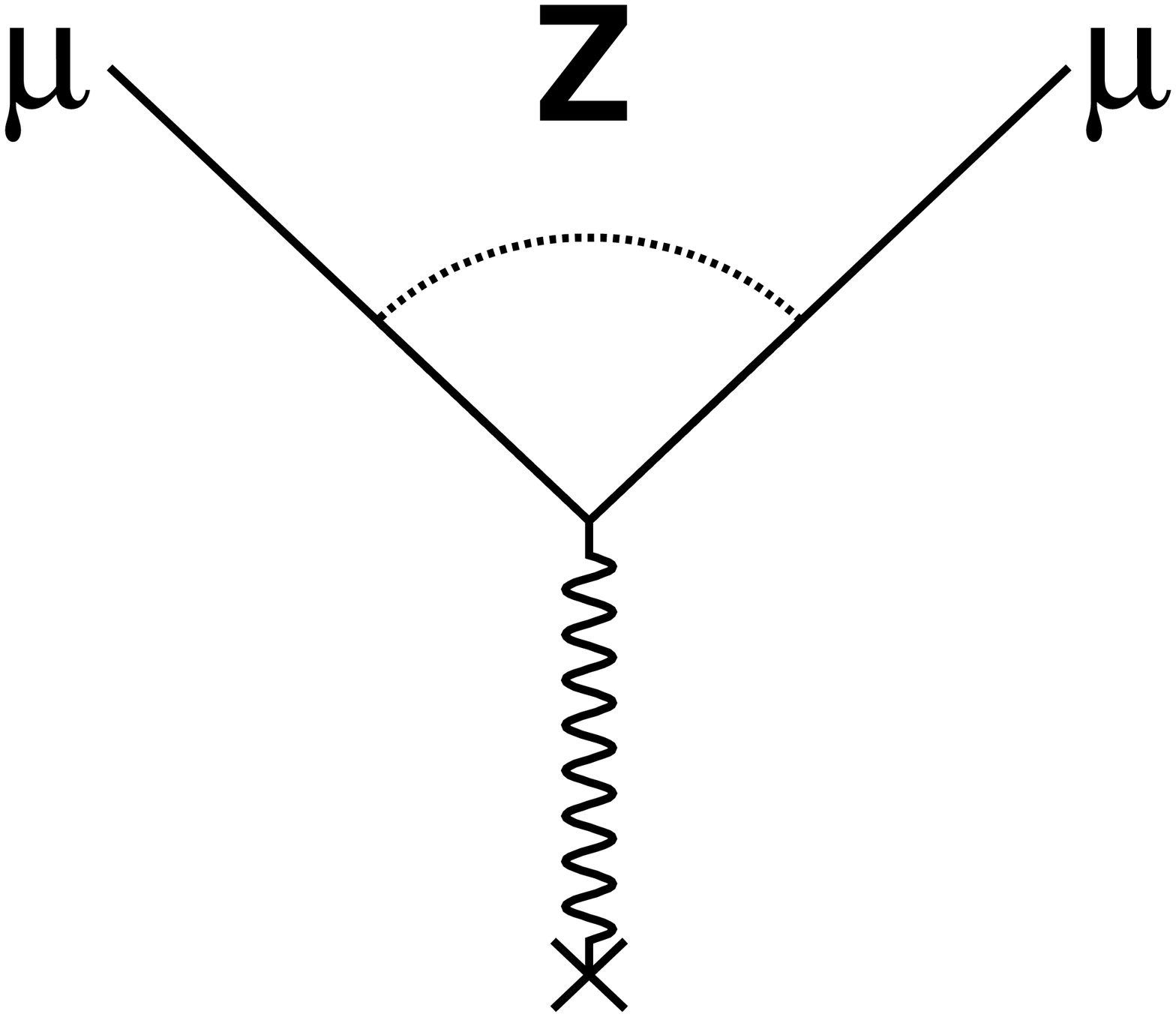,width=.24\textwidth}}
\subfigure[Hadronic]{\psfig{figure=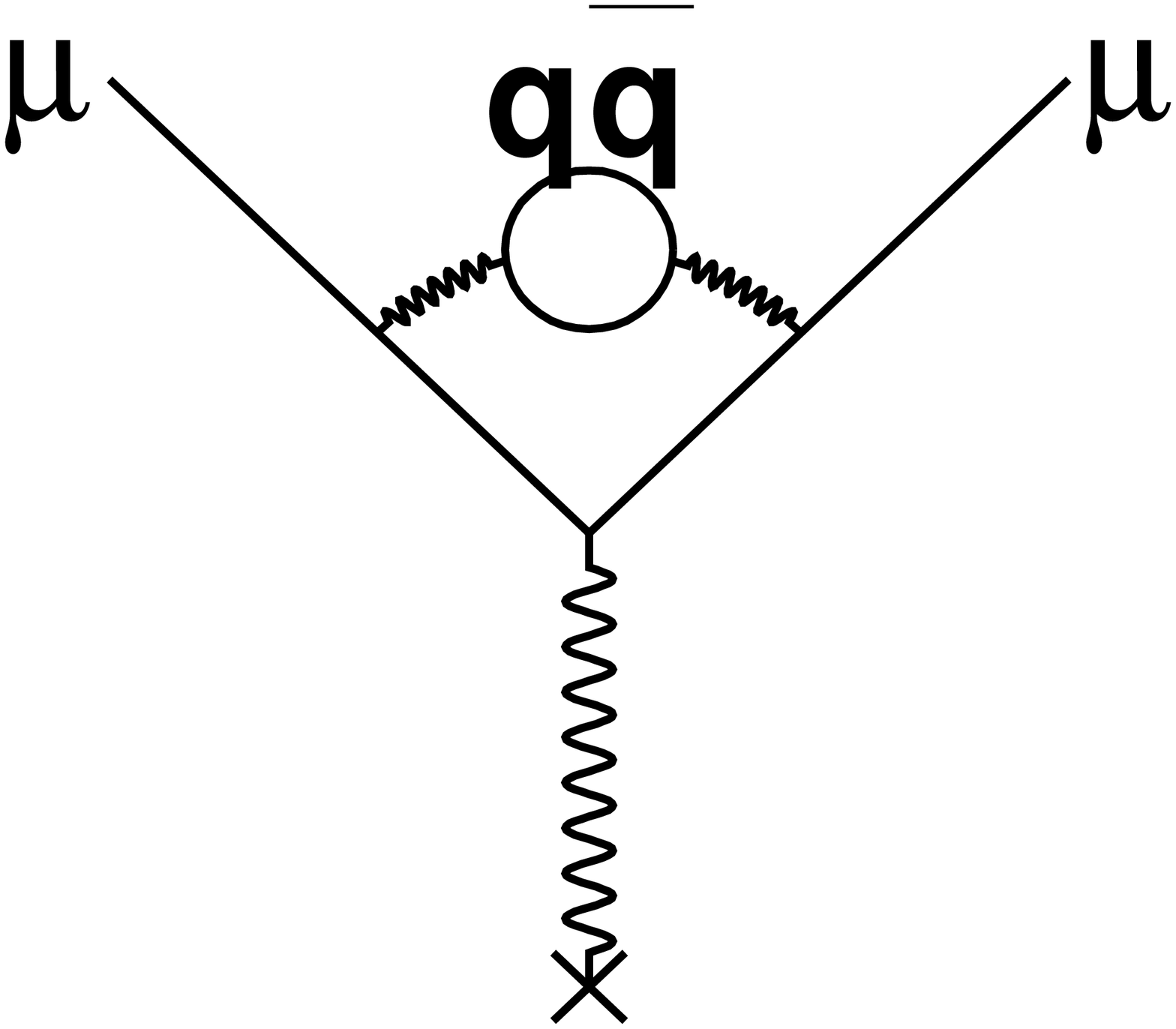,width=.24\textwidth}}
\subfigure[Light-by-light]{\psfig{figure=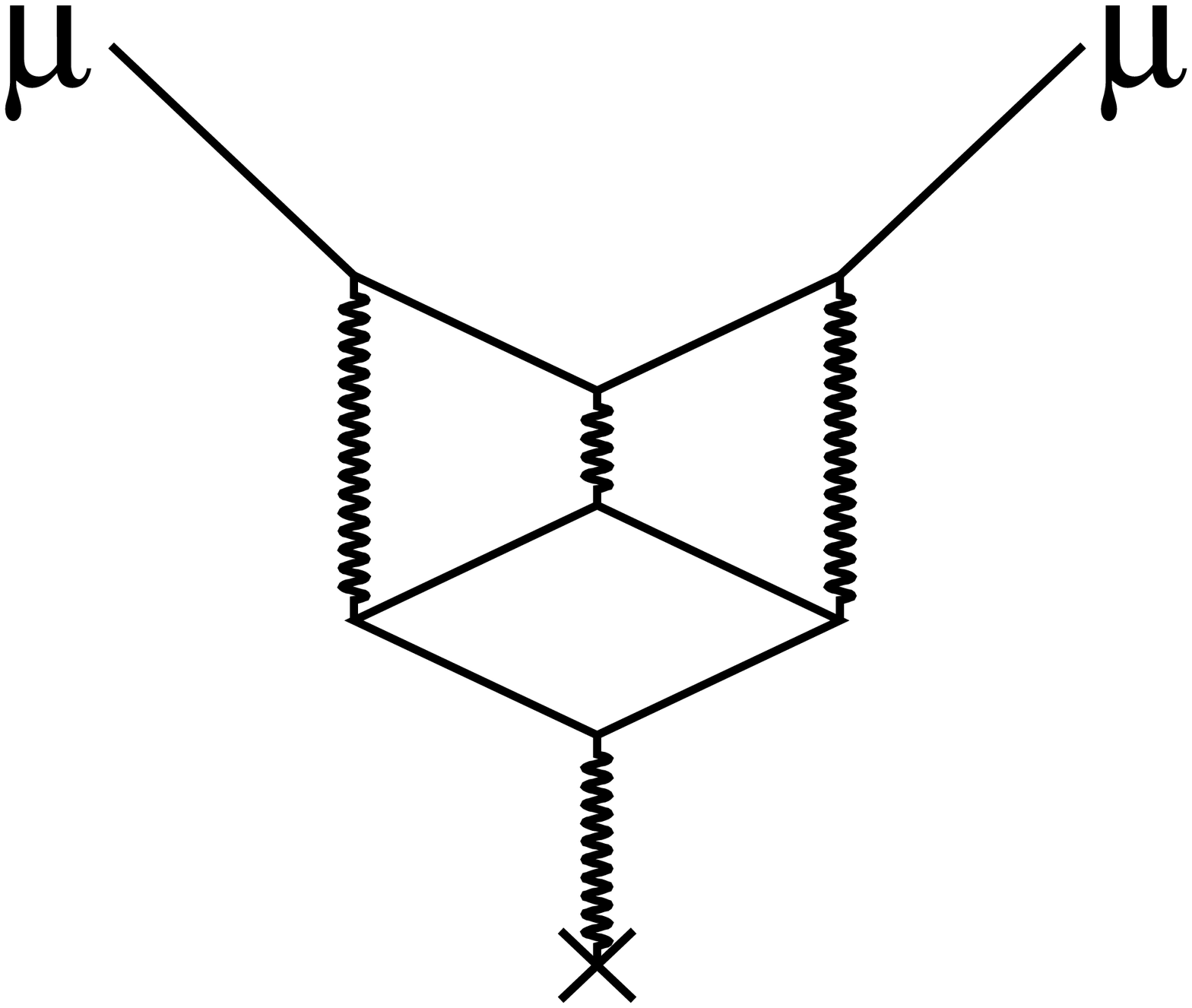,width=.24\textwidth}}
 \caption{Representative first-order Feynman diagrams for QED, weak and
 hadronic vacuum polarization, and hadronic light-by-light scattering. \label{fg:process}}
\end{center}
\end{figure}

A precision measurement of the muon anomalous magnetic moment,
$a_{\mu}~= \gm/2$, is a sensitive test of physics beyond the
standard model. Because contributions to the muon anomaly from
known processes (see Fig.~\ref{fg:process}), such as QED, the weak
interaction, and hadronic vacuum polarization (including
higher-order terms) are believed to be understood at the sub-ppm
level, any significant difference between experiment and theory
suggests a yet unknown, and thus not included, physical
process.~\cite{marciano99} Conversely, agreement between theory
and experiment can set tight constraints on new physics. Many
standard model extensions have been postulated; of these, quite a
few would manifest themselves in additional contributions to
$\amu$ at the ppm level.  This indirect method of probing
high-mass and short-distance physics is being used by the
Brookhaven E821 Collaboration~\cite{collab} in a new
measurement~\cite{carey,brown00,brown01} of \amu. The experimental
work is complemented by an aggressive effort by others to improve
the precision of the standard model theory. In the near-term
future, and generally in advance of the direct-discovery
possibilities at the Tevatron or the LHC, both experiment and
theory should achieve relative precision near or below 0.5~ppm.

The Collaboration recently reported~\cite{brown01} a 1.3~ppm
precision measurement of $\amu$ based on data obtained in the 1999
running period. A summary~\cite{marciano99} of the most up-to-date
theoretical information implied a deviation from theory by
$2.6\sigma$. Subsequently, re-examinations of the 1st-order
hadronic vacuum polarization (HVP) contribution suggested some
scatter in the central theory value, however the deviation
remained greater than two standard deviations.  Shortly after this
Workshop, Knecht and Nyffeler reported~\cite{knecht} a new
calculation of the hadronic light-by-light (LbL) diagram (see
Fig.~\ref{fg:process}d). This process cannot be determined from
measurements and therefore must be estimated using theoretical
models. It has endured a checkered past in which the {\em sign}
was once reversed from positive to negative. The new finding finds
a positive sign for the dominant pion pole contribution, but
otherwise a similar magnitude to previous studies by
others.~\cite{hayakawa98,bijnens95} This finding prompted Hayakawa
and Kinoshita to reexamine their own work~\cite{hayakawa98} which
indeed contained a sign error---in an innocent computer
algorithm---prompting a report~\cite{hayakawa01} with a new
hadronic light-by-light term of $\amu(\rm{LbL}) =
+89.6(15.4)\times 10^{-11}$. Combined with the other complete
hadronic LbL evaluation by Bijnens \ea,\cite{bijnens95} which has
also been updated to report~\cite{bijnens01} a sign error, the
standard model summary~\cite{marciano99} is \amu(\rm{thy})~$=
11~659~176.7(6.7)\times 10^{-10}$. This adjustment reduces
$\Delta\amu (\rm{exp - thy})$ by $\approx1$ standard deviation and
puts the discrepancy with the standard model at the modest $1.6
\sigma$ level.

This Workshop also featured two talks by leading contributors to
the 1st-order HVP evaluation. Simon Eidelman~\cite{eidelman}
described the recent data obtained by the Novosibirsk CMD-II team
in their experiment on $e^{+}e^{-}\rightarrow$~Hadrons in the
important $\rho$ region. These data alone, when finalized, should
reduce the uncertainty to the 0.6~ppm level without the inclusion
of hadronic tau decays. Andreas H\"ocker,\cite{hocker} together
with Michel Davier, pioneered the inclusion of hadronic tau decay
data in the evaluation of the HVP term.  Despite non-negligible
isospin breaking considerations, their results were responsible
for a significant decrease in the overall SM theory uncertainty.
But their work is not without some controversy and
questioning.\cite{melnikov} Additional tau data have been analyzed
and the present issue discussed at this Workshop was the
reliability with respect to the more direct $e^{+}e^{-}$ approach;
the question raised was, ``Can the tau data be used with
sub-percent level absolute precision?" The answer is still
unclear. On the horizon experimentally is the use of radiative
return at higher-energy $e^{+}e^{-}$ machines which promises to
add complementary precision input to the HVP database. As one has
witnessed over the past year, the standard model theory of $\amu$
is very much a work in progress, just like the experiment.

\begin{figure}
\begin{center}
\psfig{figure=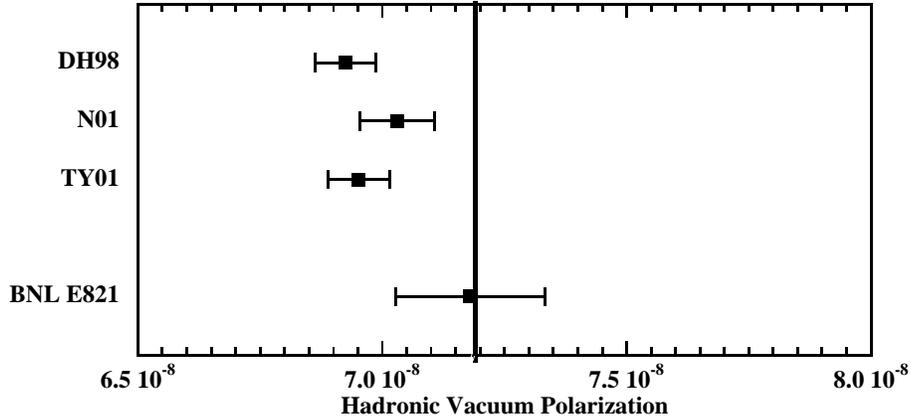,width=.75\textwidth} \caption{Recent
1st-order HVP evaluations by Davier and
H\"ocker,\protect\cite{DH98} Narison~\protect\cite{N01} and
de~Troc\'{o}niz and Yndur\'{a}in,\protect\cite{TY01} each
combining $e^{+}e^{-}$ and hadronic tau decay data. The 1999 BNL
result is shown with the QED, weak and higher-order hadronic terms
subtracted. The solid vertical line represents the current world
average experimental result. \label{fg:hadronic}}
\end{center}
\end{figure}

Recent 1st-order HVP evaluations are shown in
Fig.~\ref{fg:hadronic}. They are compared to the new E821 result
and the current world average. The QED, weak and higher-order
hadronic terms, which are believed to be known to a few tenths of
a ppm or better, have been subtracted from the measurement of
$\amu$. This results in an effective ``experiment-to-experiment"
comparison because the input to the 1st-order HVP comes from
measurement. The plot demonstrates the present relative size of
the uncertainties. Both experiment and ``theory" uncertainties
will be reduced with the analysis of additional data already
obtained.
Bill Marciano~\cite{marciano01} speculated on the implications of
the comparison of final data with settled theory. A non-standard
model result could be explained rather naturally in the context of
supersymmetry. It has been recognized for some years that the muon
anomaly scales nearly linearly with $\tan\beta$, the ratio of
Higgs doublet vacuum expectation values in the theory. The SUSY
allowed phase space certainly has ample room for large $\tan\beta$
solutions, while the lowest values, which would be exceedingly
difficult to observe in $\amu$ are beginning to be ruled out by
direct search experiments. Of course, many other candidate
explanations for non-SM $\amu$ exist; many~\cite{xxxpapers} have
been published following the original $2.6\sigma$ deviation
report.

The purpose of this paper is to describe the new experiment which
was designed following the three pioneering CERN efforts. Francis
Farley~\cite{farley} presented the historical development and the
key ideas incorporated in all $\gm$ experiments.  I will
concentrate here on the specifics related to the E821
implementation and the 1999 data analysis.

\section{The E821 Experiment}
\subsection{Principle}
The experimental goal is to measure directly \gm~ and not $g$. The
leading-order contribution to $\amu = (g-2)/2$ is the QED
``Schwinger" term whose magnitude is $(\alpha/2\pi)\approx
1.16\times 10^{-3}$. This implies that to measure $g$ directly to
an equivalent sensitivity would require an experiment with nearly
an 800-fold increase in precision.

The muon anomaly is determined from the difference between the
cyclotron and spin precession frequencies for muons contained in a
magnetic storage ring, namely
\begin{equation}
   \vec \omega_a = - {e \over m }\left[ a_{\mu} \vec B -
   \left( a_{\mu}- {1 \over \gamma^2 - 1}\right)
   \vec \beta  \times \vec E \right].
   \label{eq:omega}
\end{equation}
In principle,  an additional $\vec{\beta}\cdot\vec{B}$ term
exists. However, it vanishes when the muon trajectory is
perpendicular to the magnetic field as is the case in this
experiment.  Because electric quadrupoles are used to provide
vertical focussing in the storage ring, the $\vec{\beta}  \times
\vec{E}$
 term is necessary and illustrates the sensitivity of the spin motion to
a static electric field. This term conveniently vanishes for the
``magic" momentum of 3.094~GeV/$c$ where $\gamma = 29.3$. The
experiment is therefore built around the principle of production
and storage of muons centered at this momentum in order to
minimize the electric field effect.  Because of the finite
momentum spread of the stored muons, a modest correction to the
observed precession frequency is made to account for the muons
above or below the magic momentum. Vertical betatron oscillations
induced by the electric field imply that the plane of the muon
precession has a time-dependent pitch. Accounting for both of
these electric-field related effects introduces a $+0.81 \pm
0.08$~ppm correction to the measured precession frequency.

Muons introduced into the storage ring exhibit cyclotron motion
and their spins precess until the time of decay;
$\gamma\tau_{\mu}\approx 64.4~\mu s$.  The net spin precession
depends on the integrated path followed by a muon convoluted with
the local magnetic field experienced along the path.


Parity violation leads to a preference for the highest-energy
decay electrons to be emitted in the direction of the muon spin.
Therefore, a snapshot of the muon spin direction at time $t$ after
injection into the storage ring is obtained, again on average, by
the selection of decay electrons in the upper part of the
Lorentz-boosted Michel spectrum ($E_{max}\approx 3.1$~GeV). The
number of electrons above a selected energy threshold is modulated
at frequency $\wa$ with a threshold-dependent asymmetry
$A=A(E_{th})$. The decay electron distribution is described by
\begin{equation}
N(t) = N_{0}\exp(-t/\gamma\tau_{\mu})\left[1 + A\cos(\wa t +
\phi)\right], \label{eq:fivepar}
\end{equation}
where $N_{0}$, the normalization, $A$ and $\phi$ are all dependent
on the energy threshold $E_{th}$. For $E_{th} = 2.0$~GeV, $A
\approx 0.4$.

In summary, the experiment involves the measurement of three
quantities:  (1)~The precession frequency, $\wa$ in
Eq.~\ref{eq:omega}; (2)~The muon distribution in the storage ring;
and, (3)~The time-averaged local magnetic field.  The muon anomaly
is proportional to the ratio,
\begin{equation}
a_{\mu} \propto \frac{(1)}{\int(2)(3)}\label{eq:basic}.
\end{equation}
Term (3) above is measured using NMR in units of the free proton
precession frequency, $\omega_{p}$. Term (2) is determined from
the debunching rate of the initial beam burst and from a tracking
simulation. The combined denominator involves an event-weighted
average of the field folded with the muon distribution. The symbol
$\wpt$ represents the final average magnetic field and the muon
anomaly is computed from the expression
\begin{equation}
a_{\mu} = \frac{\wa/\wpt}{\lambda -\wa/\wpt}\label{eq:actual},
\end{equation}
where $\lambda$ is the measured~\cite{pdg} muon-to-proton magnetic
moment ratio $\mu_{\mu}/\mu_{p}=3.183~345~39(10)$.

Four independent teams evaluated $\wa$ and two studied $\wpt$.
During the analysis period, the $\wa$ and $\wpt$ teams maintained
separate, secret offsets to their measured frequencies. The
offsets were removed and $a_{\mu}$ was computed only after all
analysis checks were complete.


\section{Experiment}
\subsection{Storage Ring}
\begin{figure}
\begin{center}
\psfig{figure=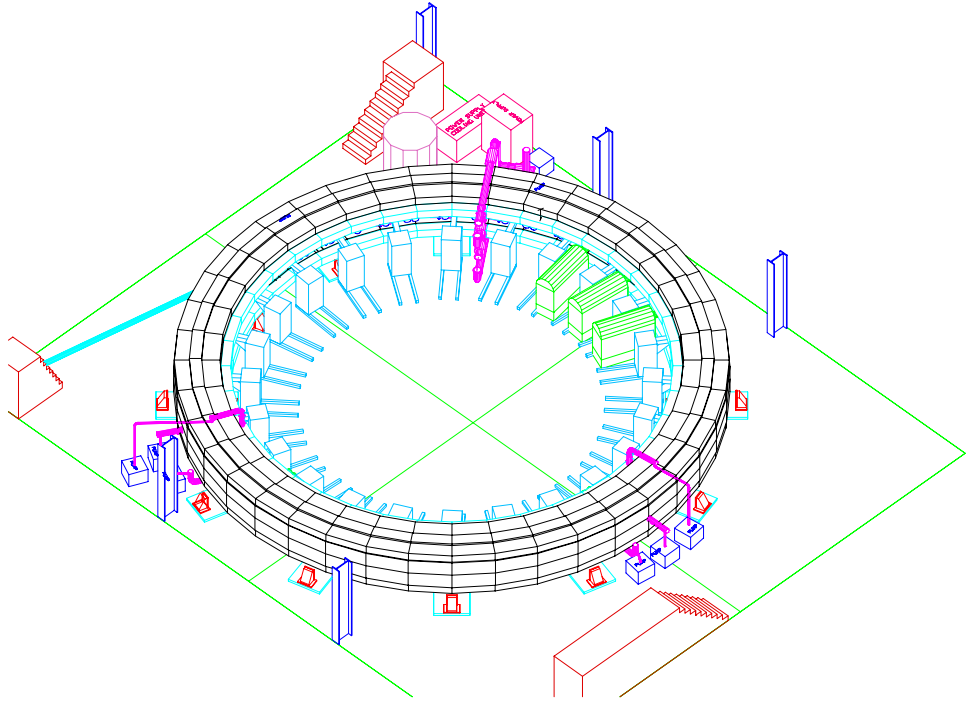,angle=90.,width=.8\textwidth} \caption{A
3D engineering rendition of the E821 muon storage ring. Muons
enter the back of the storage ring through a field-free channel at
approximately 10~o'clock in the figure.  The three modules at
approximately 2~o'clock provide the rapid current pulse which
gives the muon bunch a transverse 10~mrad kick. The regularly
spaced boxes on rails represent the electron detector systems.
\label{fg:3Dring}}
\end{center}
\end{figure}

\begin{figure}
\begin{center}
\psfig{figure=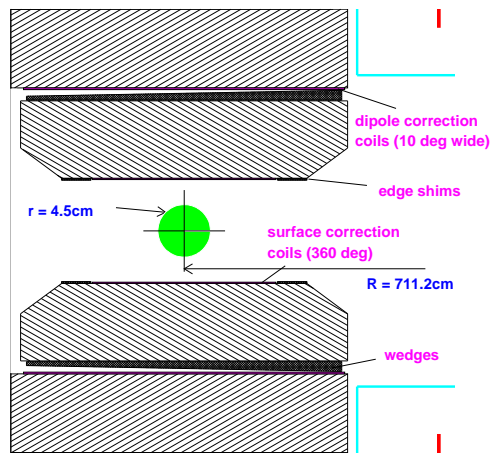,width=0.4\textwidth} \caption{Slice of
the storage ring magnet illustrating the key tools in the shimming
kit and the relative size of the storage ring volume.  The vacuum
chamber, which encloses the storage volume and houses the
quadrupoles and the trolley rails, is not shown.
\label{fg:allshims}}
\end{center}
\end{figure}

The Brookhaven muon storage ring~\cite{danby} is a superferric
``C"-shaped magnet, 7.11~m in radius, and open on the inside to
permit the decay electrons to curl inward (Fig.~\ref{fg:3Dring}).
A 5V power supply drives a 5177~A current in the three
superconducting coils. The field is designed to be vertical and
uniform at a central value of 1.4513~T. Carefully shaped steel
pole pieces and accompanying edge shims are separated from the
main iron flux return which makes up the ``C" structure.  The
field is further shaped by tapered iron wedges placed in the
yoke/pole-piece gap and by 80 low-current surface correction coils
which circumnavigate the ring on the pole piece faces. The storage
volume is 9~cm in diameter and is enclosed in a vacuum chamber. A
vertical slice through the storage ring illustrating some of these
key features is shown in Fig.~\ref{fg:allshims}.

Protons from the AGS strike a nickel target in 6--12 bunches
separated by 33~ms within a typical 2.5--3.3~s AGS cycle. Pions
created in these collisions are directed down a 72~m long beamline
to the muon storage ring. Approximately half decay,
$\pi\rightarrow \mu \nu$, and the forward going muons, having a
high degree of longitudinal polarization, remain in the channel,
generally at a slightly lower momentum. The last dipoles upstream
of the entrance to the storage ring are tuned to a momentum
$\approx 1.7\%$ lower than the main channel in order to enhance
the muon fraction in the beam; the result is a final $\mu : \pi$
ratio of approximately $50:50$ entering the ring.

A superconducting inflector magnet~\cite{krienen} provides a
field-free channel through the back of the storage ring's iron
yoke.  The field created by this magnet is tuned to cancel the
main storage ring field.  The inflector field is then prevented
from extending into the storage ring by flux trapping in a
superconducting outer shield.

The pulse structure of the AGS is translated to an effective muon
bunch of approximately 25~ns~$RMS$ passing through the exit of the
inflector into the ring. A simple circular trajectory would result
in the muon bunch striking the inflector magnet 149~ns after
injection, which is the cyclotron period. A pulsed ``kicker"
magnet provides a 10~mrad transverse deflection to the muon bunch
during the first turn in the ring. In practice, three pulsed
magnets, all in series, are made from current sheets and special
high-voltage pulsed power supplies.

The electric quadrupoles are located symmetrically at four
positions occupying, in total, $43\%$ of the ring. Immediately
after particle injection, the plates are asymmetrically charged in
order to scrape the beam against internal collimaters.  After
approximately 20~$\mu$s, the voltages are symmetrized to final
values of $\pm$24~kV which leads to the stored weak-focussing
field index $n=0.137$. The horizontal and vertical betatron
frequencies are $\omega_{x}\approx 6.23$~MHz and
$\omega_{y}\approx 2.48$MHz, respectively. Because the inflector
aperture is small compared to the cross-sectional area of the
storage volume, phase space is not fully occupied and the stored
beam exhibits ``breathing" and ``swimming" motions during a fill.
These motions manifest themselves as effective modulations in the
acceptance function of the detectors because the acceptance is
sensitive, on average, to the position of the decaying muons.  The
relevance of this fact becomes clear when one realizes that the
difference between the cyclotron and horizontal betatron
frequencies is approximately 470~kHz, which is a little greater
than twice $\wa$. These ``coherent betatron oscillations" (CBO)
are visible in the data; to account for CBO, Eq.~\ref{eq:fivepar}
must be multiplied by a term of the type
\begin{equation}
  C(t)=1 +
  A_{cbo}e^{-(t/\tau_{cbo})^2}\cos(\omega_{cbo}t+\phi_{cbo})
\label{eq:cbo}
\end{equation}
where $\tau_{cbo}$ is the time
constant for decay of the CBO terms having initial amplitude
$A_{cbo}$.

\subsection{Field Measurements}
The magnetic field is measured using pulsed nuclear magnetic
resonance (NMR) on protons in water- or Vaseline-filled probes.
The proton precession frequency is proportional to the local field
strength and is measured with respect to the same clock system
employed in the determination of \wa.  The absolute field is, in
turn, determined by comparison with a precision measurement of
$g_p$ in a spherical water sample~\cite{NMRabsolute} and is thus
determined to a precision of better than $10^{-7}$. A subset of
the 360 ``fixed" probes is used to continuously monitor the field
during data taking. Fixed probes are embedded in machined grooves
in the outer upper and lower plates of the aluminum vacuum chamber
and consequently measure the field just outside of the actual
storage volume. Constant field strength is maintained using 36 of
the  probes in a continuous feedback loop with the main magnet
power supply.

\begin{figure}
\begin{center}
\psfig{figure=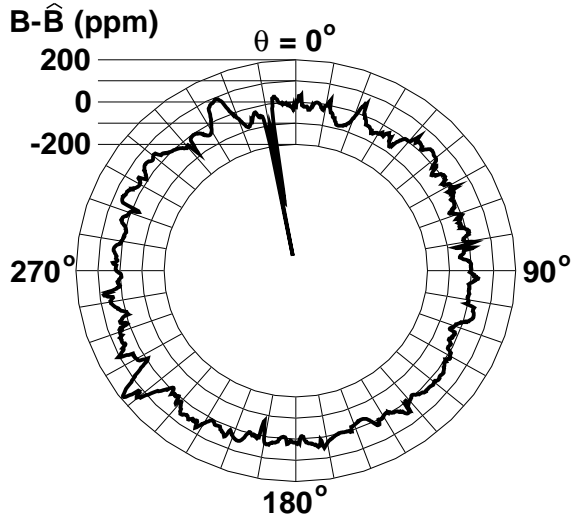,width=3in} \caption{Magnetic field
average versus azimuth from a typical trolley run.  The spike at
$350^{\circ}$ is due to a small crack in the inflector shielding
which has since been repaired. \label{fg:Gercofield}}
\end{center}
\end{figure}

The determination of the field inside the storage volume is made
by use of a unique non-magnetic trolley which can travel in vacuum
through the muon storage volume. The trolley carries 17 NMR probes
on a grid appropriate to determine the local multipolarity of the
field versus azimuth. Trolley field maps are made every few days
and take several hours to complete. The deviation from the central
field strength value $B_0 = 1.451266$~T versus azimuth is shown in
Fig.~\ref{fg:Gercofield}. At any given location, field measurement
precision is at the 0.1~ppm level even though the local field
varies by as much as 100~ppm from $B_0$. However, it is the {\em
average} field which must be determined to the sub~ppm level and
this goal is readily achieved. A 800~ppm ``spike" in the field is
seen in a $1^{\circ}$ azimuthal segment. This small defect is due
to a crack in the inflector magnet shielding. Special measurements
were made to map the field in this region for the 1999 data; this
inflector was replaced for the 2000 and 2001 data taking periods.
A one-time systematic error of 0.2~ppm is included in the 1999
result to account for uncertainties in the procedure of measuring
the errant field in this region.

\begin{figure}
\begin{center}
\psfig{figure=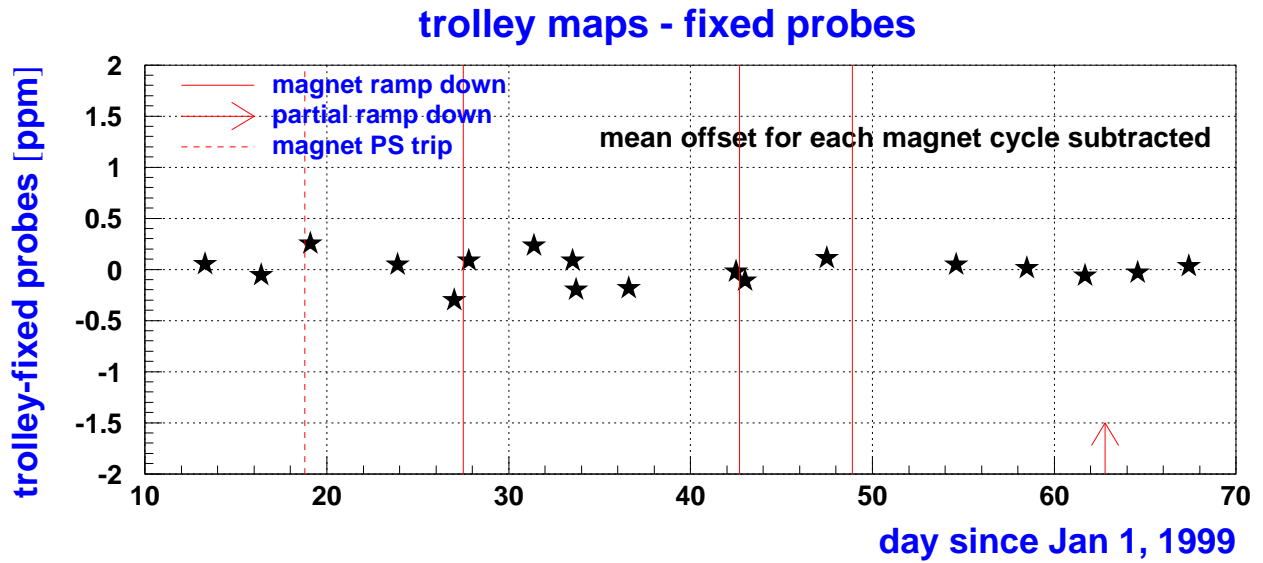,width=\textwidth} \caption{The
magnetic field strength as measured at the center of the storage
ring using the NMR trolley system compared to that which was
expected from data obtained using the fixed NMR probes. Each point
represents an individual trolly run during the 1999 data taking
period. The agreement is at the 0.15~ppm level.
\label{fg:tracking}}
\end{center}
\end{figure}

Comparison of the azimuthally averaged trolley field with that
which is computed using  the fixed probe data permits a determine
of the time-dependent field. Figure~\ref{fg:tracking} illustrates
the excellent agreement between these two systems. The relative
field is interpolated from the fixed-probe extrapolation to better
than 0.15~ppm.

\begin{figure}
\begin{center}
\psfig{figure=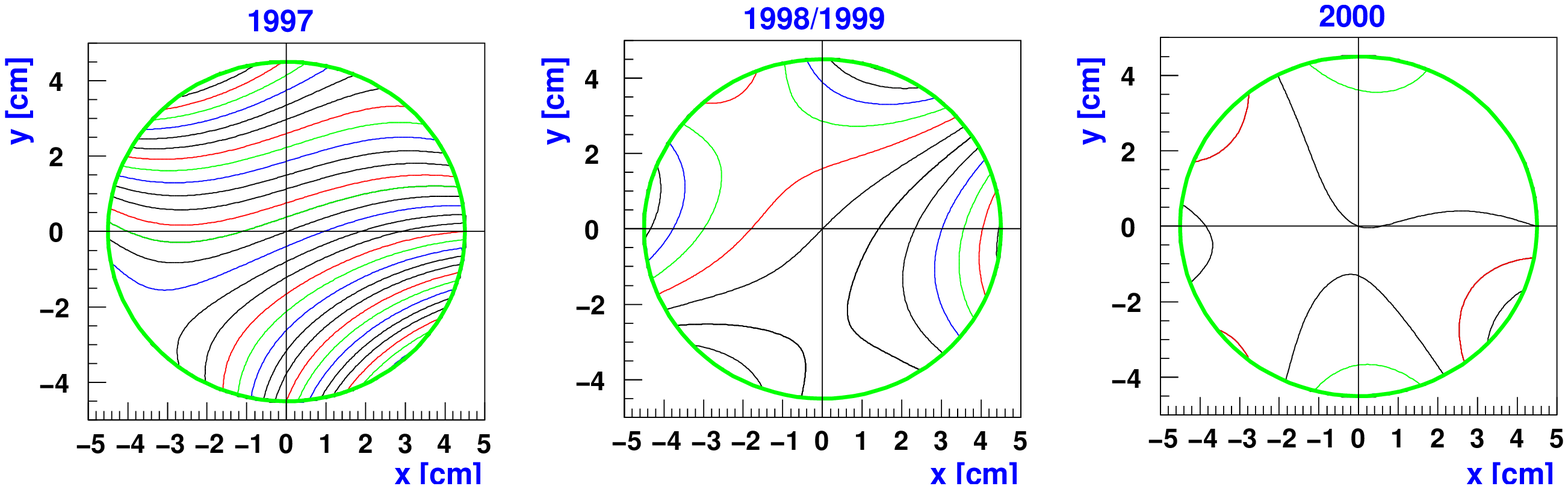,width=\textwidth}
\caption{Sequence of improving magnetic field profiles, averaged
over azimuth and
           interpolated using a multipole expansion. The
           circle indicates the storage aperture. From left to right,
           these maps represent the field variations for the 1997
           commissioning run, the 1998/1999 data taking, and finally
           the 2000 run.  Improvements in shimming and replacement of
           the inflector magnet led to the two major improvements
           following the 1997 and 1999 runs, respectively.
\label{fg:fieldmaps}}
\end{center}
\end{figure}

The trolley measurements also establish the field homogeneity
throughout the storage volume. Three snapshots of the
azimuthally-averaged field are shown in Fig.~\ref{fg:fieldmaps}.
The contours represent 1~ppm deviations from the central value.
The left panel illustrates the relatively uneven field achieved
shortly after the ring was commissioned in 1997; here it is
already better than the CERN~III field.\cite{cernIII}  An
aggressive shimming program led to the improved contours for the
1998 and 1999 data taking periods (middle panel). With the
inflector replaced, the field for the 2000 and 2001 data taking
was again significantly improved (right panel).

\section{Electron Detection and \wa\ Analysis}
\begin{figure}
\begin{center}
\psfig{figure=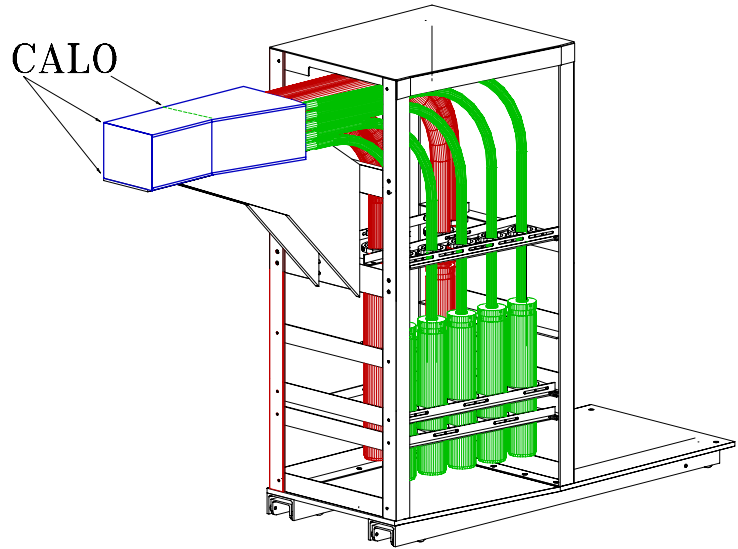,angle=90,width=.6\textwidth}
\caption{Electromagnetic calorimeter system for the experiment.
The figure illustrates the four lightguides from the calorimeter
and the five which couple to the scintillator hodoscope.
\label{fg:calo}}
\end{center}
\end{figure}

\begin{figure}
\begin{center}
\psfig{figure=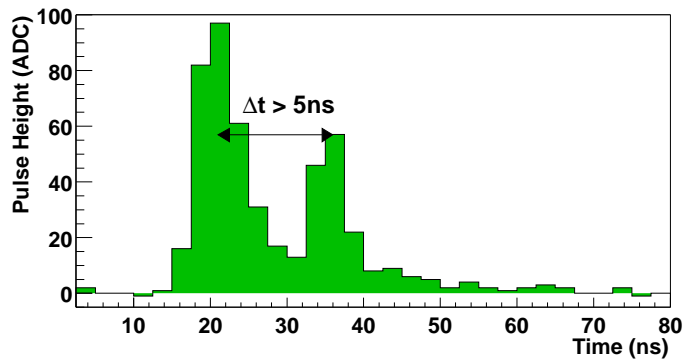,width=.6\textwidth} \caption{An
``island" of waveform digitizer samples.  In this representation,
the constant offset has been subtracted.  Two well-separated
pulses are shown.  The pulse-finding routines have no difficulty
with such events.  Special care is needed when the second pulse
moves within 5~ns of the first event. \label{fg:samples}}
\end{center}
\end{figure}

The electron detection system consists of 24 lead-scintillating
fiber electromagnetic calorimeters~\cite{sedykh} located
symmetrically around the inside of the storage ring and placed
immediately adjacent to the vacuum chamber.  The 23~cm long,
radially-oriented fiber grid terminates on four lightguides which
pipe the light to independent Hamamatsu R1828 2-inch PMTs, see
Fig.~\ref{fg:calo}.  The PMT gains are carefully balanced because
the four analog signals are added prior to sampling by a custom
400~MHz waveform digitizer.  At least 16 digitized samples
(usually 24 or more) are recorded for each decay electron event
exceeding a hardware threshold of approximately 1~GeV. An example
of a series of such samples is shown in Fig.~\ref{fg:samples}.
Offline pulse-finding and fitting establishes electron energy and
muon time of decay.

\begin{figure}
\begin{center}
\psfig{figure=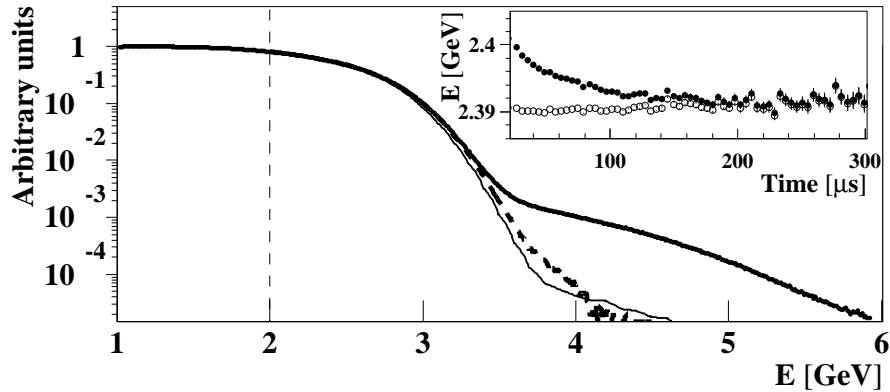,width=.75\textwidth} \caption{Energy
spectrum at late times only (thin line), at all times (thick line)
and after pileup subtraction (dashed line).  The inset compares
the average energy in a full \gm~ cycle versus time for
unsubtracted (solid circles) and subtracted (open circles)
spectra. The energy distribution must be flat.
\label{fg:pileupenergy}}
\end{center}
\end{figure}

Because the full raw waveforms are stored, multi-particle pileup
can be studied with sophisticated fitting algorithms which are
superior to any ``online," hardware discriminator techniques. The
approach taken is to fit waveforms for all occurring pulses using
known detector response functions. We identify the level of pileup
and remove such events, on average.  By processing the data using
different artificial deadtime windows around or near the driving
trigger pulse, and by carefully accounting for the energy found in
such windows, a pileup-free spectrum (less than $10\%$ pileup
remaining) can be created by appropriate, normalized difference
spectra. Figure~\ref{fg:pileupenergy} illustrates three curves of
the energy spectrum in the calorimeters. The solid curve is the
``true" energy distribution found at late times where no pileup
events are expected. The dashed distribution, which extends to
much higher energies, is observed at early times when the rate in
the detectors is high. Pileup produces effective electron energies
beyond the natural maximum of 3.1~GeV. The dotted curve is the
recreated pileup-free final spectrum. Its congruence with the
late-time curve gives confidence in the correction procedure.

\begin{figure}
\begin{center}
\psfig{figure=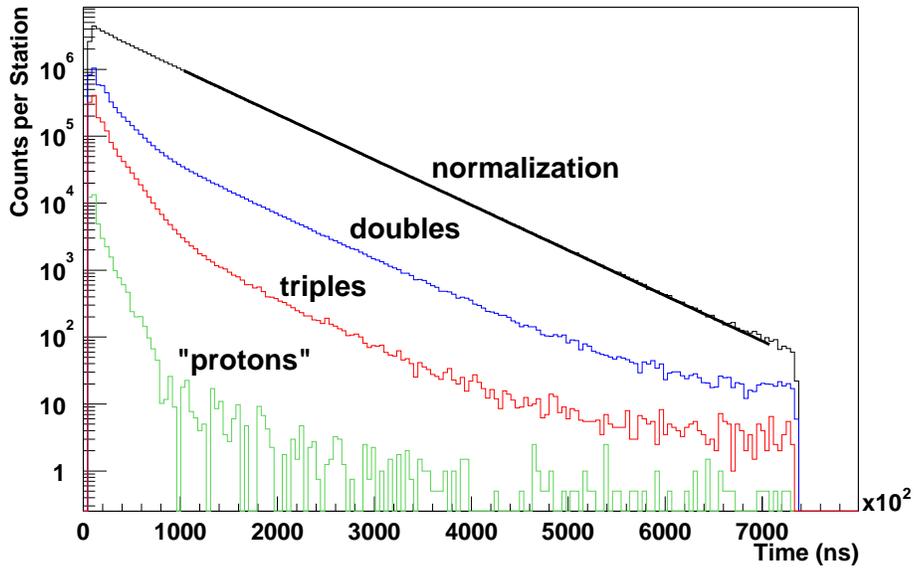,angle=-90,width=.75\textwidth}
\caption{A muon loss distribution from one detector (1999 data).
Binning in this histogram is in units of the $\gm$ period, thus
the normalization curve is an exponential with lifetime of
approximately 64~$\mu$s.  The triples curve is proportional to the
muon loss rate. After $\approx 100~\mu$s, the loss rate is
essentially constant. At earlier times, the loss function is
described by an exponential with a lifetime of approximately
$27~\mu$s. The doubles curve is similar but, with a lower
coincidence requirement, contains background.  The ``proton" curve
involves three FSD coincidences and significant energy in the
calorimeter of the third station. This occurs from occasional
hadronic showers. The triples and doubles curves also show proton
events at late times ($\approx 700 \mu$s) when most muons have
decayed. \label{fg:muonloss}}
\end{center}
\end{figure}

Five-fold, vertically segmented scintillator hodoscopes (FSDs) are
attached to the front face of many of the calorimeters. These FSDs
are used to measure the rate of ``lost muons" from the storage
ring. During the initial scraping period, and extending for a
period of time afterwards, errant muon trajectories will pass
through one of the aperature-defining collimaters inside the
storage ring. These muons lose enough energy to exit the ring,
many following a path which penetrates three consecutive
calorimeters and their corresponding FSDs. A triple coincidence of
FSDs, together with the absence of significant energy in the
calorimeters, is proportional to the lost muon rate. A
representative muon loss spectra for one detector is shown in
Fig.~\ref{fg:muonloss} At early times, the loss rate exceeds the
natural muon decay rate and can be described by an exponential
with an approximate $27~\mu s$ lifetime. At later times, the loss
rate is essentially constant.  The overall loss rate is
proportional to
\begin{equation}
    \Lambda(t)=1+A_\Lambda e^{-t/\tau_\Lambda}.
   \label{eq:muonloss}
\end{equation}
Integration of $\Lambda(t)$ and inclusion in the fitting function
is necessary to obtain a good representation of the instantaneous
muon population in the ring.

\section{Fitting the 1999 Data}
\begin{figure}
\begin{center}
\psfig{figure=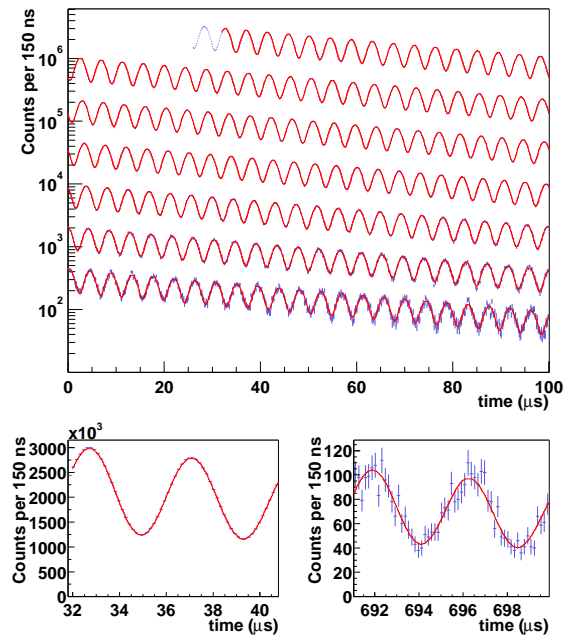,width=.5\textwidth} \caption{The 1999
data and a simple 5-parameter fit.  While the fit looks good, see
the lower panels, the $\chi^2$ is too high. A more complete
fitting function, applied to a pileup-subtracted data set, yielded
$\chi^2$ consistent with 1. \label{fg:wiggles}}
\end{center}
\end{figure}

The five adjustable parameters in Eq.~\ref{eq:fivepar} were
sufficient to give an excellent $\chi^2$ when used to fit the 1998
data.\cite{brown00} Figure~\ref{fg:wiggles} illustrates data and
5-parameter fit for 1999.  By eye, the fit also looks excellent;
however, the $\chi^2$ is unacceptable because the CBO, pileup and
muon losses have not been accounted for in this 20 times larger
data set. The ``pileup-free" spectrum is fit to the function
\begin{equation}
    f(t)=N(t) \cdot C(t) \cdot \Lambda(t)
   \label{eq:finalfit}
\end{equation}
where $N(t)$ is the ideal function (Eq.~\ref{eq:fivepar}), $C(t)$
accounts for the CBO (Eq.~\ref{eq:cbo}), and $\Lambda(t)$
characterizes the time-dependent muon lossses
(Eq.~\ref{eq:muonloss}).  With Eq.~\ref{eq:finalfit}, or
equivalent fitting strategies used by the other analysis teams,
$\chi^2$ is consistent with 1.

\begin{figure}
\begin{center}
\psfig{figure=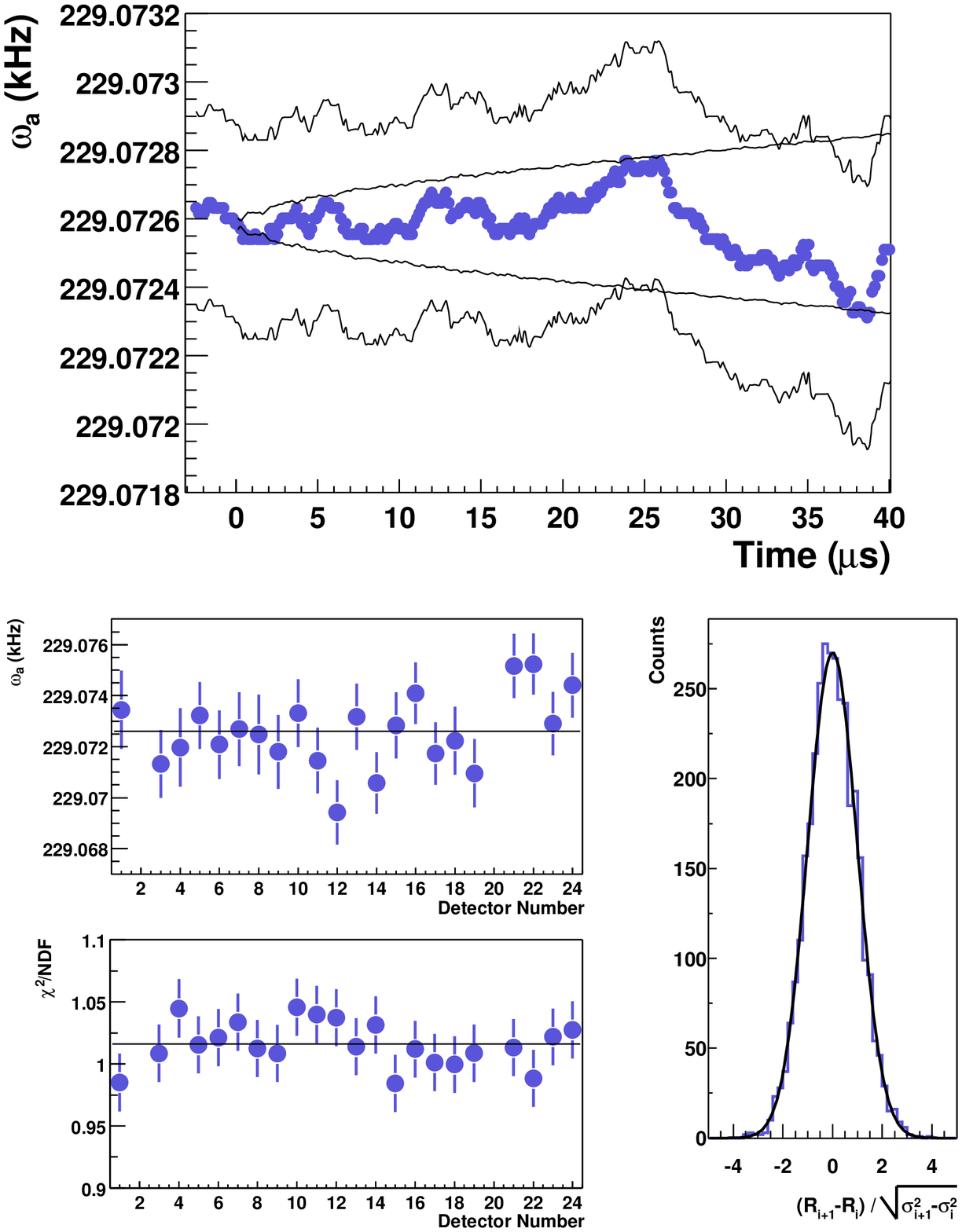,width=\textwidth} \caption{A
series of ``consistency" plots from one of the $\wa$ analyses.
The top plot shows the stability of the fit for $\wa$ with respect
to the start time of the fit.  The solid bands represent the
allowed $1\sigma$ deviation from the first fit point based on the
understanding that later fits are highly correlated with the
starting fit. The plot on the bottom right is the deviation of fit
results for adjacent start times.  It should be centered at zero
with a width of 1 for deviations which are purely statistical.
The panels on the lower left illustrate the stability of the fit
versus detector number and the $\chi^2$ for each individual
detector fit. \label{fg:fitresults}}
\end{center}
\end{figure}

Deviations  from the ideal function of the type described are more
significant at early times compared to late times.  One commonly
used test employed by the analysis teams is to fit the data
progressively beginning at different ``start"  times.  As shown in
Fig.~\ref{fg:fitresults}, $\wa$ only fluctuates within the
expected band but is otherwise stable. The distribution of
variances from neighboring fits is shown on the lower right panel
of the figure. This plot has a mean consistent with zero and a
normalized width consistent with 1, as expected. The bottom left
panels of the figure give $\wa$ and the $\chi^2$ by detector for
the fixed start time of 30~$\mu$s. These and other statistical
tests yield satisfactory results.

\section{Results, Conclusions and Outlook}
Table~\ref{table:results} summarizes the results. The experimental
value from 1999 for positive muons is $a_{\mu^+}~=
11\,659\,203(15)\times10^{-10}$ (1.3~ppm.  With the updated
standard model theory value of
$\amu(\rm{thy})=11~659~176.7(6.7)\times 10^{-10}$ (0.57~ppm), one
finds $\Delta a_{\mu^+}(\rm{exp-thy})=25(16) \times 10^{-10}$. The
2000 $\mu^+$ data is presently being analyzed and should reduce
the uncertainty considerably. The 2001 $\mu^-$ data is just
beginning to be processed at the time of this writing.  To
complete the experiment, with equal positive and negative muon
event samples, a final negative muon run will be necessary. The
Collaboration is eager to complete this data taking phase and
accompanying analysis in order to achieve our ultimate goal of
0.4~ppm precision or better in the measurement of the muon
anomaly.

\begin{table}
\begin{center}
\begin{tabular}{|l|l|c|}
\hline Item & Value & Relative Uncertainty \\ \hline
$\wpt/2\pi$ & $61~791~256(25)$~Hz & 0.4 ppm \\
$\wa/2\pi$ & $229~072.8(3)$~Hz & 1.3 ppm \\
Electric field correction & +0.81 ppm & $10\%$ \\
Combined $\wpt$ systematic uncertainties & --- & 0.4 ppm \\
Combined $\wa$ systematic uncertainties & --- & 0.3 ppm \\
Final experimental value & $11\,659\,203(15)\times10^{-10}$ & 1.3 ppm \\
Updated SM theory & $11~659~176.7(6.7)\times 10^{-10}$ & 0.57 ppm
\\
$\Delta(a_{\mu^+}(\rm{exp-thy})$ & $25(16)\times 10^{-10}$ &
$1.6~\sigma$
\\ \hline
\end{tabular}
\caption{Result summary for the 1999 data
analysis.}\label{table:results}
\end{center}
\end{table}

\section*{Acknowledgments}
The \gm~ experiment is supported in part by the U.S. Department of
Energy, the U.S. National Science Foundation, the German
Bundesminister f\"{u}r Bildung und Forschung, the Russian Minsitry
of Science, and the US-Japan Agreement in High Energy Physics. The
author would like to thank the organizers of the Workshop at Erice
for an excellent meeting.

\end{document}